\documentstyle[prd,aps,eqsecnum,floats]{revtex}
\draft
\input epsf

\begin{document}
\preprint{PU-RCG/99-8, gr-qc/990826}

\renewcommand{\topfraction}{0.99}
\renewcommand{\bottomfraction}{0.99}
\twocolumn[\hsize\textwidth\columnwidth\hsize\csname 
@twocolumnfalse\endcsname

\title{Self-similar cosmological solutions with a non-minimally coupled
scalar field}
\author{Damien J.~Holden and David Wands}
\address{School of Computer Science and Mathematics, University of
Portsmouth, Portsmouth PO1 2EG,~~~U.~K.}
\date{\today}
\maketitle
\begin{abstract}
We present self-similar cosmological solutions for a barotropic fluid
plus scalar field with Brans-Dicke-type coupling to the spacetime
curvature and an arbitrary power-law potential energy.  We identify
all the fixed points in the autonomous phase-plane, including a
scaling solution where the fluid density scales with the scalar
field's kinetic and potential energy. This is related by a conformal
transformation to a scaling solution for a scalar field with
exponential potential minimally coupled to the spacetime curvature,
but non-minimally coupled to the barotropic fluid. Radiation is
automatically decoupled from the scalar field, but energy transfer
between the field and non-relativistic dark matter can lead to a
change to an accelerated expansion at late times in the Einstein
frame. The scalar field density can mimic a cosmological constant even
for steep potentials in the strong coupling limit.
\end{abstract}

\pacs{PACS numbers: 98.80.Cq, 04.50.+h 
\hfill
Preprint PU-RCG/99-8, gr-qc/9908026\\
{\em to appear in Physical Review D}
}

\vskip2pc]


\section{Introduction}

Scalar fields have come to play a dominant role in recent years in
theoretical models of the universe. This has usually been in the
context of inflationary models of the very early universe where the
self interaction potential energy density $V(\phi)$ remains undiluted
by the cosmological expansion. If the potential is sufficiently flat
this can lead to an effective cosmological constant which can drive an
accelerated expansion. The detailed nature of the evolution is driven
by the specific form of the scalar field's potential energy.

More recently observational evidence has been growing that the energy
density of the universe today may be dominated by a homogeneous
(smooth) component with negative pressure - christened
``quintessence''~\cite{Triangle}.  A suitably small cosmological
constant $\Lambda$ provides a good fit to the observational
data~\cite{Tegmark,EfsBriLas99}, but the required low energy scale,
$\Lambda \sim (10^{-3}{\rm eV})^4$, seems unnatural in any
supergravity model.  This has lead to an explosion of interest in the
late-time evolution of self-interacting scalar fields, where radiation
and matter contribute the dominant energy density at earlier times,
but slowly rolling scalar fields may provide a dynamical mechanism to
achieve a small effective cosmological constant today by mimicking a
decaying effective-$\Lambda$.  The late time evolution of sufficiently
steep potentials has been shown to be largely insensitive to initial
conditions~\cite{St98} (though it should be remembered that it the case
of a true cosmological constant there is also no dependence upon
initial conditions!). This steep potential must become flat by the
present in order to act as an effective cosmological constant today
and there remains the ``fine-tuning'' problem of why the scalar field
density comes to dominate at just the present epoch.

Scalar fields with exponential potentials provide a strikingly simple
model in which the energy density of the scalar field decreases at
exactly the same rate as a barotropic fluid.
Wetterich~\cite{Wetterich88} was the first to point out that the
energy density of a scalar field with an exponential potential does not
simply die away if the potential is too steep to drive inflation, but
rather approaches an attractor solution~\cite{CopLidWan98} where it
mimics the equation of state of a barotropic fluid (radiation or dust)
and remains a fixed fraction of the total energy density.  We will
refer to these solutions as {\em self-similar} or {\em scaling}\footnote
{Note that the term ``scaling solution'' has been applied elsewhere in the
literature~\cite{LidSch99} to describe any power-law evolution for the
scalar field density in a radiation or matter dominated background.}
cosmological models where the dimensionless density parameter for each
component, $\Omega_i\equiv \rho_i/\rho_{\rm crit}$, is a constant.

Self-similar homogeneous cosmological solutions are invariant
under a global conformal rescaling of the metric. Such models
may be expanding, but the physical state at different times
differ only by a change in the overall length scale~\cite{WaiEll97}.
In a Friedmann-Robertson-Walker (FRW) cosmological model this is
equivalent to a rescaling of the cosmic time,
\begin{equation}
\label{scalet}
t\to \Gamma t \ ,
\end{equation}
where $\Gamma$ is a constant. Self-similarity requires that the
evolution of the scale factor $a(t)$ is a power-law, and the Hubble
constant $H\equiv \dot{a}/a\propto 1/t$.

We will consider the evolution of spatially flat FRW cosmologies
containing a fluid, with density $\rho$, and pressure $P$, and a
scalar field $\phi$ with self-interaction potential $V(\phi)$.
In general relativity, the Friedmann constraint equation requires
\begin{equation}
H^2 = {8\pi G_N \over 3} \left( \rho + {1\over2} \dot\phi^2 + V(\phi)
\right) \ ,
\end{equation}
where $G_N$ is Newton's constant. Equation~(\ref{scalet}) then
requires that the matter content is invariant under a rescaling
$\rho\to\Gamma^{-2}\rho$. Requiring $P\to\Gamma^{-2}P$ for all
$\Gamma$ then leads to a barotropic equation of state
$P=(\gamma-1)\rho$ where $\gamma$ is a constant. Thus the familiar
radiation ($\gamma=4/3$) or pressureless matter ($\gamma=1$) dominated FRW
models can be described as self-similar solutions with
\begin{equation}
\label{abaro}
a \propto t^{2/(3\gamma)} \ .
\end{equation}
A scalar field has kinetic energy $\dot\phi^2/2$ which only allows a
constant shift of the scalar field, $\phi\to\phi+\phi_\Gamma$, if we
are to require $\dot\phi^2\to\Gamma^{-2}\dot\phi^2$. In order to
obtain a self-similar solution we require in addition
\begin{equation}
V(\phi) \to V(\phi+\phi_\Gamma)
 = \Gamma^{-2}V(\phi) \ .
\end{equation}
which is only compatible with a non-interacting field, $V(\phi)=0$, or
an exponential potential
\begin{equation}
V(\phi) = V_0 \exp \left( -\lambda\kappa\phi \right) \ ,
\end{equation}
where $\kappa^2=8\pi G_N$, $\lambda$ is a dimensionless constant and
we have $\kappa\phi_\Gamma=(2/\lambda)\ln\Gamma$.

FRW solutions for scalar fields with exponential potentials where
the kinetic energy and potential energy of the field remain
proportional were proposed as a model for power-law inflation in
the early universe by Luchinn and Matarrese~\cite{LM85}, and are
the late-time attractor solutions (in the absence of other matter)
for $\lambda^2<6$~\cite{Halliwell87,BB88}.  More recently
attention has focused on the possible late-time evolution of
scalar fields in FRW cosmologies containing matter. Self-similar
solutions are known for scalar fields with exponential potentials
whose energy density scales with that of a barotropic fluid
yielding the same time dependence given in Eq.~(\ref{abaro}) for a
barotropic fluid dominated solution~\cite{Wetterich88}. These
scaling solutions are the unique late-time attractors for
sufficiently steep potentials
($\lambda^2>3\gamma$)~\cite{CopLidWan98}. Such a scalar field is
so successful at scaling with the barotropic matter that the
scalar field never comes to dominate the cosmological dynamics. A
minimally coupled field can only mimic a cosmological constant at
late times if the scale invariance of the scalar field is broken,
leading to the deviation from the pure scaling solution which must
be fine-tuned to occur at just the present epoch.

In this paper we will consider the effect of non-minimal coupling
of the scalar field to the spacetime curvature. In the next
section we show that self-similar cosmological solutions are
possible for scalar fields with simple power-law potentials if the
scalar field has a Brans-Dicke-type coupling to the spacetime
curvature.  The equations of motion reduce to an autonomous
phase-plane whose fixed points correspond to self-similar
solutions. We show that this system is conformally related to a
scalar field with exponential potential and minimally coupled to
the spacetime curvature, but with an explicit coupling to matter,
studied in Refs.~\cite{Wands93,Wetterich95,Amendola99,Bill99,Bill99thesis}. 
This coupling is
sensitive to the fluid equation of state and can produce a natural
change in the nature of the solution after matter-radiation
equality.  In contrast with the case of a minimally coupled scalar
field there is the possibility of recovering an effective late
time cosmological constant even if the potential remains steep~\cite{texas},
depending on the value of the dimensionless non--minimal coupling
constant and we investigate this possibility in
section~\ref{SectQ}.

\section{Brans--Dicke--type cosmology with power--law potential}

It is possible to obtain self-similar solutions for scalar fields with
arbitrary power-law potentials , $V(\phi)=V_0(\kappa\phi)^{2n}$, if we
go beyond Einstein's theory of general relativity and allow the field
to be non-minimally coupled to the spacetime curvature.
In Brans-Dicke gravity~\cite{BraDic61} with dimensionless parameter
$\omega$, Newton's constant is replaced by a dynamical field
$G=\omega/(2\pi\phi^2)$
and the generalised Friedmann equation requires
\begin{equation}
H^2 = {8\pi \over 3} {\omega \over 2\pi\phi^2}
 \left( \rho + {1\over2} \dot\phi^2 +
 V(\phi) - {3\over2\omega}H\phi\dot\phi \right) \ .
\end{equation}
In the original Brans-Dicke theory where $V(\phi)=0$,
Eq.~(\ref{scalet}) is compatible with a global re-scaling\footnote
{The global rescaling leaves the dimensionless Brans-Dicke parameter,
$\omega$, invariant. A local rescaling $\Gamma(\phi)$ rescales the
Brans-Dicke parameter~\cite{LidWanCop99}.}
of the barotropic fluid density $\rho\to\Gamma_\rho\rho$
and $\phi\to\Gamma_\phi\phi$ for all
$\Gamma_\rho/\Gamma_\phi^2=\Gamma^{-2}$. The self-similar
solution for a barotropic fluid in Brans-Dicke gravity was given by
Nariai~\cite{Nariai68,HolWan98}.
In the presence of a power-law potential, $V(\phi)=V_0(\kappa\phi)^{2n}$,
(absent in the pure Brans-Dicke theory~\cite{BraDic61}) we
require in addition that
$V(\phi)/\phi^2\to\Gamma^{-2}[V(\phi)/\phi^2]$ which specifies
$\phi\to\Gamma^{-1/(n-1)}\phi$ and $\rho\to\Gamma^{-2n/(n-1)}\rho$.
Thus self-similar solutions can exist for ``quintessence''-type power-law
scalar field potentials~\cite{RatPee88,Quintessence}, so long as this scalar
field has Brans-Dicke type coupling to the spacetime curvature.

Motivated by the preceding qualitative discussion of self-similar
solutions we examine the cosmological evolution of one of the simplest
forms of non--minimal coupling proposed by Brans and
Dicke~\cite{BraDic61} which is described by a single dimensionless
parameter $\omega$.  In addition to the original Brans--Dicke
Lagrangian~\cite{BraDic61}, we introduce a self interaction potential
$V(\phi)$~\cite{Bergmann68,Wagoner70,Nordtvedt70,BarMae90}.
The action is
\begin{eqnarray}
\label{action} S & = & \int \sqrt{-g} d^4x [ \pm {1\over8\omega}
\phi^2 R \mp \frac{1}{2} g^{\mu\nu} \phi_{,\mu}\phi_{,\nu} -
V(\phi)] \, \nonumber \\ & + & \int \sqrt{-g}d^4x {\cal
L}_{matter} \,
\end{eqnarray}
where upper/lower signs should be chosen to ensure $\pm\omega>0$ and
hence a positive gravitational coupling.
The self-interaction potential is taken to be a power-law,
\begin{equation}
\label{defVphi}
V(\phi) = V_0 (\kappa\phi)^{2n}\ .
\end{equation}

This action can be re-expressed as a theory of interacting matter
fields in general relativity with fixed Newton's constant
$G_N$~\cite{Dicke62} if we define quantities in the conformally related
Einstein frame with respect to the rescaled metric
\begin{equation}
\tilde{g}_{\mu\nu} =
\left| {\kappa^2\phi^2\over 4\omega} \right| g_{\mu\nu}\ .
\end{equation}
The scalar field is now minimally coupled to metric
$\tilde{g}_{\mu\nu}$, but non-minimally coupled to the other matter
fields.  We will re-express the scalar field $\phi$ in terms of a
field $\psi$ which has a canonical kinetic term in the Einstein frame,
which requires that
\begin{equation}
\kappa \phi = \exp \left( {\kappa\psi \over \sqrt{2(3+2\omega)}}
\right) \ .
\end{equation}
The scalar field will have a non-negative energy density in the
Einstein frame so long as $\omega>-3/2$. Thus we consider $\omega$
with any value $>-3/2$, which includes, for example, the coupling of
the string-theory dilaton where $\omega=-1$~\cite{string}.

The effective self-interaction potential for canonical field
$\psi$ in the Einstein frame is of exponential form and given
by (see, e.g., Refs.~\cite{BarMae90,Bill98})
\begin{equation}
\tilde{V}(\psi) = \left( \frac{4\omega}{\kappa^2\phi^2} \right)^2 V(\phi)
= \tilde{V}_0\exp(-\lambda\kappa\psi) \ ,
\end{equation}
where $\tilde{V}_0 \equiv (4\omega)^2V_0$ and
\begin{equation}
\label{deflambda}
\lambda \equiv \sqrt{ {2\over3+2\omega} } (2-n) \ .
\end{equation}

\subsection{Autonomous Phase--Plane}

Using the Hubble rate, $\tilde{H}$, fluid density, $\tilde{\rho}$, and
scalar field density,
$\tilde\rho_\psi=(1/2)(d\psi/d\tilde{t})^2+\tilde{V}(\psi)$, defined
in the Einstein frame, the evolution equations are
\begin{eqnarray}
\label{evolH} \frac{d\tilde{H}}{d\tilde{t}}  = -\frac{\kappa^2}{2}
\left(\tilde{\rho} + \tilde{P} +
\left(\frac{d\psi}{d\tilde{t}}\right)^2 \right) \,,
\\
\label{evolrho} \frac{d\tilde{\rho}}{d\tilde{t}} +
3\tilde{H}(\tilde{\rho} + \tilde{P})  =  - \frac{dQ}{d\tilde{t}}
\,,
\\
\label{evolpsi} \frac{d\psi}{d\tilde{t}} \left(
\frac{d^2\psi}{d\tilde{t}^2} + \frac{d\tilde{V}}{d\psi} +
3\tilde{H}\frac{d\psi}{d\tilde{t}} \right)  =
\frac{dQ}{d\tilde{t}} \,,
\end{eqnarray}
subject to the Friedmann constraint,
\begin{equation}
\tilde{H}^2 = \frac{\kappa^2}{3}\left(\tilde{\rho} +
\frac{1}{2} \left(\frac{d\psi}{d\tilde{t}}\right)^2 + \tilde{V}(\psi)
\right).
\end{equation}
The non-minimal coupling of the scalar field in the Brans-Dicke frame
leads to an explicit energy transfer in the Einstein frame,
between the scalar field and the fluid,
\begin{equation}
\label{defQ}
\frac{dQ}{d\tilde{t}} \equiv
 \kappa \frac{d\psi}{d\tilde{t}}
 \frac{\tilde\rho-3\tilde{P}}{\sqrt{2(3+2\omega)}}
\,.
\end{equation}
The same system of equations was considered, starting from a
different motivation, by Wetterich~\cite{Wetterich95}, and more
recently by Amendola~\cite{Amendola99}. In the case
$dQ/d\tilde{t}=0$, these equations reduce to previous studies of
minimally coupled scalar fields with exponential
potentials~\cite{Wetterich88,CopLidWan98}.

Following Ref.~\cite{CopLidWan98}, we define,
\begin{equation}
x \equiv
 \frac{\kappa}{\sqrt{6}\tilde{H}} \left( \frac{d\psi}{d\tilde{t}} \right)
\qquad , \qquad
y \equiv
 \frac{\kappa\sqrt{\tilde{V}}}{\sqrt{3}\tilde{H}} \,.
\end{equation}
The evolution equations for a barotropic fluid where
$\tilde{P}=(\gamma-1)\tilde\rho$ can then be written
as~\cite{CopLidWan98,Amendola99}
\begin{eqnarray}
\label{xprime} x' & = & -3x + \lambda \sqrt{\frac{3}{2}}y^2 +
\frac{3}{2}x [ 2x^2 +\gamma(1-x^2-y^2)] \, \nonumber \\ & + &
W(1-x^2-y^2),
\\
\label{yprime} y' & = & -\lambda \sqrt{\frac{3}{2}}xy +
\frac{3}{2}y [ 2x^2 +\gamma(1-x^2-y^2)],
\end{eqnarray}
where a prime denotes the derivative with respect to $N\equiv \int
\tilde{H}d\tilde{t}$, the logarithm of the scale factor in the Einstein frame,
and we have used the constraint equation,
\begin{equation}
\label{xyconstraint}
\frac{\kappa^2\tilde{\rho}}{3\tilde{H}^2} +x^2 +y^2 = 1
\end{equation}

We parameterise the energy transfer in terms of the dimensionless
quantity $W$, where
\begin{equation}
\label{defW}
W \equiv \sqrt{\frac{3}{3+2\omega}}\left(\frac{4-3\gamma}{2}\right)\ ,
\end{equation}
Equations~(\ref{xprime}) and~(\ref{yprime}) define an two-dimensional
autonomous phase-plane whenever $W$ can be written as function of $x$
and $y$. The Brans-Dicke-type theory defined in Eq.~(\ref{action})
naturally leads to an interaction with constant $W$, although the
analysis could be extended to more general 
$W(x,y)$~\cite{Bill99,Bill99thesis}.

The constraint Eq.~(\ref{xyconstraint}) restricts physical solutions
with non--negative fluid density to $0 \le x^2+y^2 \le 1$ and so the
evolution is completely described by trajectories within the unit
disc.  In the following discussion we will only consider expanding
cosmologies corresponding to the upper-half disc, $y\ge0$.

\subsection{Self-similar solutions}
\label{sect2.2}

Fixed points of the system, ($x_i$,$y_i$) where $x'=0$ and $y'=0$,
correspond to power-law solutions for the scale factor
and logarithmic evolution of the scalar field with respect to the
cosmic time in the Einstein frame:
\begin{equation}
\label{power}
\tilde{a} \propto \tilde{t}^{\tilde{p}} \quad , \quad \kappa\psi \propto
\tilde{q}\ln\tilde{t} \,,
\end{equation}
where the constants $\tilde{p}$ and $\tilde{q}$ are given by,
\begin{eqnarray}
\tilde{p} & = & \frac{2}{3\gamma(1-x_i^2-y_i^2) + 6x_i^2}\,,
\nonumber \\ \tilde{q} & = &
\frac{2\sqrt{6}x_i}{3\gamma(1-x_i^2-y_i^2) + 6x_i^2} \,.
\end{eqnarray}
These are self-similar solutions where the dimensionless density
parameter for the barotropic fluid in the Einstein frame, given from
the constraint Eq.~(\ref{xyconstraint})
\begin{equation}
\tilde\Omega \equiv {\kappa^2\tilde\rho \over 3 \tilde{H}^2} = 1 - x_i^2 -
y_i^2 \ ,
\end{equation}
is a constant and therefore so is the dimensionless density of the
scalar field
\begin{equation}
\label{defOmegapsi}
\tilde\Omega_\psi=x_i^2+y_i^2 \ .
\end{equation}
The scalar field has an
effective barotropic index given by,
\begin{equation}
\label{defgammapsi}
\tilde\gamma_\psi \equiv
\frac{\tilde{P}_\psi+\tilde{\rho}_\psi}{\tilde{\rho}_\psi} =
\frac{2x^2}{x^2+y^2}.
\end{equation}

The existence and stability of critical points, already defined
in the Einstein frame, remain unchanged under a conformal
transformation back to the Brans-Dicke frame. The evolutionary
behaviour for the scale factor and scalar field is however
modified. If we denote the evolutionary behaviour of the scale factor
$a$, and scalar field $\phi$ in the Brans-Dicke frame by,
\begin{equation}
a \sim t^p \qquad , \qquad \phi^2 \sim t^q
\end{equation}
exponents can be related to those in the Einstein frame by,
\begin{eqnarray}
p & = & \tilde{p} +
\frac{(\tilde{p}-1)\tilde{q}}{\sqrt{2(3+2\omega)}-\tilde{q}}\,,
\nonumber \\ q  & = &
\frac{2\tilde{q}}{\sqrt{2(3+2\omega)}-\tilde{q}}\,.
\end{eqnarray}

Depending on the values of the parameters $\lambda, \gamma$ and $W$ we
can have up to five fixed points in the Einstein frame. The nature and
stability of each point in the phase-plane is briefly reviewed
below. A full analysis of the existence and stability is given in the
Appendix. See also Ref.~\cite{Amendola99}.

\subsubsection{2--Way, Matter--Kinetic scaling solution}

\begin{equation}
x_1 = \frac{2W}{3(2-\gamma)} \qquad , \qquad  y_1=0.
\end{equation}
This solution lies on the $x$--axis where the scalar field potential is
negligible, and the scalar field's density in the Einstein frame is
dominated by its kinetic energy, leading to a stiff equation of state
for the scalar field, $\tilde\gamma_\psi=2$.
We have a power-law solution of the form given in Eq.~(\ref{power}) with

\begin{eqnarray}
\tilde{p}_1 & = & \frac{6(2-\gamma)}{9\gamma(2-\gamma) + 4W^2}\,,
\nonumber \\ \tilde{q}_1 & = & \frac{4\sqrt{6}W}{9\gamma(2-\gamma)
+ 4W^2}\,.
\end{eqnarray}

In the Brans--Dicke frame these are the power law solutions for a barotropic
fluid in Brans-Dicke gravity as first given by Nariai~\cite{Nariai68,HolWan98},
where the power law exponents of the scale factor and
scalar field are given by
\begin{eqnarray}
p_1 & = & \frac{2\omega(2-\gamma) + 2}{3\omega\gamma(2-\gamma) +
4}\, \nonumber \\ q_1 & = &
\frac{2(4-3\gamma)}{3\omega\gamma(2-\gamma) + 4}\,.
\end{eqnarray}

\subsubsection{Kinetic dominated solutions}

\begin{equation}
x_{2,3} = \pm 1 \qquad , \qquad  y_{2,3}=0
\end{equation}
These solutions correspond to negligible fluid density and negligible
scalar field potential so that the Friedmann constraint equation is
dominated by the kinetic energy of the scalar field with a stiff
equation of state $\tilde{\gamma}_\psi=2$.
We have power-law solutions of the form given in Eq.~(\ref{power}) with
\begin{equation}
\tilde{p}_{2,3} = 1/3 \qquad , \qquad \tilde{q}_{2,3} =
\pm\sqrt{2/3} \,.
\end{equation}

In the Brans--Dicke frame we recover the vacuum solutions of Brans-Dicke
gravity found by O'Hanlon and Tupper~\cite{OHT72,HolWan98} with power-law
exponents given by
\begin{eqnarray}
p_{2,3} & = & \frac{\sqrt{6} \mp \sqrt{2(3+2\omega)}}{\sqrt{6} \mp
3\sqrt{2(3+2\omega)}}\,, \nonumber \\ q_{2,3} & = & \frac{\pm
2\sqrt{6}}{3\sqrt{2(3+2\omega)} \mp \sqrt{6}}\,.
\end{eqnarray}

\subsubsection{Scalar field dominated solutions}

\begin{equation}
x_4 = \lambda/\sqrt{6} \qquad , \qquad y_4 = (1-\lambda^2/6)^{1/2}.
\end{equation}
Here the fluid density is negligible, but neither the kinetic energy,
nor the potential dominates the energy density of the scalar field in
the Einstein frame. The scalar field has an effective equation of state
$\tilde{\gamma}_\psi=\lambda^2/3$, and for $\lambda^2<2$ this solution
corresponds to power-law inflation~\cite{LM85}.
For $\lambda\neq0$ we have power-law solutions of the form given in
Eq.~(\ref{power}) with
\begin{equation}
\label{tildepq4}
\tilde{p}_4 = \frac{2}{\lambda^2} \qquad
, \qquad
\tilde{q}_4 = \frac{2}{\lambda} \,.
\end{equation}

In the Brans--Dicke frame the power law exponents are given (for
$n\neq1$ and $n\neq2$) by~\cite{BarMae90},
\begin{eqnarray}
p_4 & = & {2\omega+n+1 \over (2-n)(1-n)}\,, \nonumber \\ q_4  & =
& {2 \over 1-n} \,.
\end{eqnarray}
For $n=0$ the potential acts like a false vacuum energy density in the
Brans-Dicke frame and this corresponds to extended
inflation solutions for $\omega>1/2$~\cite{extinf}.
For $n=2$ it is the potential in the Einstein frame that remains
constant, leading to de Sitter expansion~\cite{BarMae90,SanGre97}.
The case $n=1$ was studied by Kolitch~\cite{Kolitch96}.

\subsubsection{3--Way, Matter--Kinetic--Potential scaling solutions}

\begin{eqnarray}
\label{x5y5}
x_5 & = & \frac{3\gamma}{\sqrt{6}\lambda-2W}\,,
\nonumber \\ y_5^2 & = & \left[ \frac{9\gamma(2-\gamma) -
2W(\sqrt{6}\lambda-2W)}{(\sqrt{6}\lambda-2W)^2}\right].
\end{eqnarray}
Here neither the fluid nor the scalar field dominates the evolution, and
we have self-similar solution where both the potential and kinetic
energy density of the scalar field remains proportional to that of the
barotropic matter. The effective equation of state for the scalar field
is given by,
\begin{equation}
\tilde\gamma_{\psi5} = \gamma \left(
\frac{6\gamma}{6\gamma-W(\sqrt{6}\lambda-2W)} \right) \ .
\end{equation}
We have power-law solutions of the form given in Eq.~(\ref{power})
with~\cite{Wetterich95,Amendola99}
\begin{equation}
\label{tildepq5}
\tilde{p}_5 = \frac{2}{3\gamma}\left(
\frac{\sqrt{6}\lambda-2W}{\sqrt{6}\lambda}\right) \qquad , \qquad
\tilde{q}_5 = \frac{2}{\lambda}.
\end{equation}

In the Brans-Dicke frame this corresponds to power-law evolution
with exponents given by,
\begin{equation}
p_5 =  \frac{2}{3\gamma} \left( \frac{n}{n-1} \right) \qquad ,
\qquad q_5 = {2 \over 1-n} \ .
\end{equation}
As far as we are aware, this solution has not been discussed before in
the context of Brans-Dicke gravity.  It is interesting to note that
the cosmological evolution in the Brans-Dicke frame is independent of
the Brans-Dicke parameter $\omega$, although it does determine the
existence of this 3-way scaling solution (see Appendix).

\subsection{Recovering the Einstein limit when $\omega\to\infty$}

There are two limiting procedures which one can apply to recover
minimally coupled general relativistic solutions.

One can work with quantities defined with respect to the Einstein
frame where we always have a fixed gravitational coupling $G_N$. We
can take the limit $W\to0$ to decouple the barotropic fluid from the
scalar field but keep $\lambda$ finite. {}From Eqs.~(\ref{deflambda})
and~(\ref{defW}) this requires $\omega\to\infty$ but also
$|n|\to\infty$ in the Brans-Dicke frame. In this way we obtain a
minimally coupled scalar field in Einstein gravity with an exponential
self-interaction potential which possesses self-similar solutions
where the energy density of the field scales with that of the
barotropic matter for $\lambda^2>3\gamma$~\cite{Wetterich88}.  However
in this case the scalar field never changes the cosmological dynamics
from that in a conventional universe dominated by the barotropic fluid
on its own, so these solutions by themselves can never produce an
accelerating universe at late times.

Alternatively one can let $\omega\to\infty$ in the Brans-Dicke frame
while keeping $G=\omega/(2\pi\phi^2)$, $V(\phi)$ and $n$ finite. This
means that finite changes in the scalar field $\phi\to\phi+\Delta\phi$
leave both the gravitational coupling $G$ and the potential energy $V$
fixed. This corresponds to the familiar general
relativistic solution for a massless scalar field plus cosmological
constant.  The cosmological constant will always dominate over any
barotropic fluid with $\gamma>0$, but there is always a fine-tuning
required if we are to postpone this domination until the present
epoch.

In the context of a scalar-tensor gravity theory it is intriguing to
note the implicit connection between fixing the gravitational coupling
and fixing the effective value of the cosmological constant. A theory
in which $\omega$ has small effective value in the early universe can
naturally give rise to a self-similar cosmological evolution where the
potential energy scales with radiation or dust. If $\omega$ itself can
dynamically evolve to large values and fix the gravitational coupling
at late times~\cite{DamNor93} then it would also fix the value of the
effective cosmological constant. But a variable $\omega(\phi)$ theory
breaks the scale-invariance of the pure Brans-Dicke theory and
requiring that $\phi$ becomes fixed only at late-cosmological times is
another manifestation of the cosmological coincidence
problem~\cite{St98}.

\section{Cosmological quintessence}
\label{SectQ}

We will now consider whether the self-similar solutions presented
above could form the basis of a viable cosmological model which
produces an effective cosmological constant dominating only at
late-times without fine-tuning.  There are two key features which
appear promising in this respect.  Firstly the scalar field is
automatically decoupled, $W=0$, for relativistic matter (radiation)
with a traceless energy-momentum tensor, $\gamma=4/3$. In this case
the Einstein frame evolution coincides with that for a minimally
coupled scalar field with exponential potential which has scaling
solutions for $\lambda^2>3\gamma$. Secondly, we naturally expect a
change in the cosmological evolution when non-relativistic
(pressureless) matter with $\gamma=1$ comes to dominate the density at
late times. Here the non-minimal coupling could play a key role and,
at least in principle, it seems possible to effect a change to an
accelerating universe.

These two phases of evolution are fundamentally different and the
change from radiation to matter domination (``matter-radiation
equality'') is known to occur relatively late in the cosmological
history. We consider separately these two phases of cosmological
evolution.

In a multi-fluid cosmology we need to consider the possibility that
different fluids are minimally coupled with respect to different
metrics. Relativistic matter (radiation) is minimally coupled in any
conformally related frame, but in the case of non-relativistic matter
we will distinguish two alternative scenarios:
\begin{enumerate}
\item
{\em Brans-Dicke theory:} In the original gravity theory proposed by
Brans and Dicke~\cite{BraDic61} it assumed that all matter was
minimally coupled with respect to the Brans-Dicke frame metric
$g_{\mu\nu}$. This respects the weak equivalence
principle~\cite{Will93}. Thus all matter, ordinary baryonic matter
and any dark matter, must be minimally coupled to scalar field, $\phi$,
in the Brans-Dicke frame. Conformally transforming to the Einstein
frame introduces an explicit energy transfer to the scalar field,
$\psi$, [given by $dQ/d\tilde{t}$ in Eq.~(\ref{defQ})] for all
non-relativistic matter.
\item
{\em Generalised dark matter:} Given the uncertain form of the cold
dark matter assumed to dominate the matter density in the universe
today, we will also consider the possibility that this dark matter and
ordinary baryonic matter are minimally coupled with respect to
different metrics~\cite{DamGibGun90,CasGar92,Wands93}. This violates the weak
equivalence principle but experimental tests~\cite{Will93} are only
performed with baryonic matter and baryonic observers. We will
consider in what follows the case where baryonic matter is minimally
coupled in the Einstein frame ($dQ_{\rm B}/d\tilde{t}=0$). This is
equivalent to setting $\omega_{\rm B}\to\infty$ as far as baryonic
observers are concerned, which is consistent with weak-field
tests~\cite{Will93}. But some or all of the dark matter is minimally
coupled in a Brans-Dicke frame with finite $\omega_{\rm DM}$ which
leads to $dQ_{\rm DM}/d\tilde{t}\neq0$ in the Einstein frame.
\end{enumerate}

Amendola~\cite{Amendola99} considered case (1) where observational
constraints on the value of $\omega$ restrict the cosmological
evolution to be close to the conventional general relativistic
case. Instead we will consider the second possibility of generalised
dark matter where solar system tests of weak-field gravity do not
directly constrain $\omega_{\rm DM}$. It is important that in this
case cosmological observations of baryonic matter refer to the
quantities calculated in the Einstein frame.

\subsection{Radiation}

The late-time attractor for a decoupled barotropic fluid ($W=0$)
plus scalar field with an exponential potential in the Einstein
frame is either the scalar field dominated solution given in
Eq.~(\ref{tildepq4}) for flat potentials with $\lambda^2<3\gamma$
or the 3-way scaling solution given in Eq.~(\ref{tildepq5}) for
steeper potentials with $\lambda^2>3\gamma$. This latter solution
yields $\tilde{a} \propto \tilde{t}^{\tilde{p}}$ where
$\tilde{p}=2/(3\gamma)$ as previously
studied~\cite{Wetterich88,CopLidWan98}.

The energy-momentum tensor for radiation with $\gamma=4/3$ has
vanishing trace, so it is automatically decoupled from a scalar field
non-minimally coupled to the spacetime scalar curvature, as shown by
Eq.~(\ref{defW}). The only constraint that can be placed upon the
evolution of the field during this radiation era is that the scalar
field should not disrupt the standard temperature-time relation at
nucleosynthesis. If ordinary barotropic matter, specifically protons
and neutrons, as well as electrons and neutrinos are minimally coupled
in the Einstein frame, then this constraint must be applied in this
frame where $\tilde{p}=1/2$ is unperturbed, and we simply require that
the scalar field does not contribute a significant fraction of the
overall energy-density. Substituting Eq.~(\ref{x5y5}) for the 3-way
scaling solution into Eq.~(\ref{defOmegapsi}) we have
\begin{equation}
\label{nuc}
\lambda^2 > \frac{4}{\tilde{\Omega}_{\psi,{\rm nuc}}^{\rm Max}} \ ,
\end{equation}
where the current bound on the energy density of the scalar field at
the time of nucleosynthesis, $\tilde{\Omega}_{\psi,nuc}^{Max}$, is in
the range 0.03 to 0.2~\cite{FerJoy98,Bur99}.

For large $|\lambda|$ the steepness of the effective potential,
ensures, rather paradoxically, that the evolution of the scalar field
$\psi$ is {\em slow}. This is because the gradient of an exponential
potential depends exponentially upon the value of the scalar field. As
the scaling solution is approached the gradient of the scalar field is
balanced against the frictional effect of the radiation density
Eq.~(\ref{evolpsi}). The larger the exponent, the slower the scalar
field must evolve in order to maintain scaling.  {}From
Eq.~(\ref{deflambda}) we see that the limit $|\lambda|\to\infty$ can
be obtained either as $|n|\to\infty$ for $\omega>-3/2$ or for finite
$n$ when $\omega\to-3/2$. This is in marked contrast to the more
familiar case of nucleosynthesis constraints on Brans-Dicke gravity
where we require weak-coupling
$\omega\to\infty$~\cite{DamGun91,CasGarQui92}.

\subsection{Non-relativistic matter}

Most estimates of the present density of non-relativistic (pressureless)
matter in the universe are in the range $\Omega_{m,0} =
0.2-0.4$~\cite{Turner98,Fre99}. There are upper limits on the
fraction of baryonic matter coming from models of primordial
nucleosynthesis. Observational upper bounds on the abundance of
deuterium in the universe places an upper limit on the baryonic
matter content, $\Omega_{\rm B}h^2 = 0.02$~\cite{Bur99},
leaving most of the matter as some unknown cold dark matter.
We consider here the cosmological evolution where this CDM is
non-minimally coupled to the scalar field $\psi$ in the Einstein frame.

In addition recent observations imply that at the present time the
universe is undergoing an accelerated expansion
\cite{Garnavich98,Perlmutter98}, which equates to $\tilde{p}>1$ in the
Einstein frame for power-law solutions presented in
Sect.~\ref{sect2.2}. This is only possible for the scalar field
dominated solution given in Eq.~(\ref{tildepq4}) if $\lambda^2<2$,
which would lead to acceleration even during the radiation era which is
wildly incompatible with primordial nucleosynthesis, or for the 3--way
matter--kinetic--potential scaling solution given in
Eq.~(\ref{tildepq5}) for $\lambda<0$.  In practice the nucleosynthesis
constraint on $\lambda^2$ during the radiation era, given in
Eq.~(\ref{nuc}), ensures that the 3-way scaling solution must
exist and is the unique late-time attractor for non-relativistic
matter ($\gamma=1$) and $\lambda<0$.

Acceleration, $\tilde{p}_5>1$, for the 3--way matter--kinetic--potential
scaling solution with $\gamma=1$ and $\lambda<0$ requires
\begin{equation}
\label{lowerboundW}
 -\sqrt{6}\lambda < 4W \ .
\end{equation}
which from Eqs.~(\ref{deflambda}) and~(\ref{defW}) implies that
$2<n<3$ in the Brans-Dicke field potential in Eq.~(\ref{defVphi}).
The existence condition (see Appendix) reduces to
\begin{equation}
-\sqrt{6}\lambda > - W + \sqrt{W^2+18} \ ,
\end{equation}
which, combined with Eq.~(\ref{lowerboundW}) gives as a necessary
condition for an accelerated 3-way scaling solution that
$W>\sqrt{3}/2$ which, from Eq.~(\ref{defW}), implies $\omega<1/2$.
Because nucleosynthesis places such a strong lower limit on the allowed
value of $\lambda^2$ this requires the coupling constant $\omega$, to
be close to its minimum value of $-3/2$ for $2<n<3$.

The present day density parameter for the scalar field $\psi$ is
then given from Eqs.~(\ref{defOmegapsi}) and~(\ref{x5y5}) by,
\begin{eqnarray}
\tilde{\Omega}_\psi & = & {2W(2W-\sqrt{6}\lambda)+18 \over
(2W-\sqrt{6}\lambda)^2}\,, \nonumber \\ & = & {2n+15+12\omega_{\rm
DM} \over (2n-3)^2}\,.
\end{eqnarray}
Requiring an accelerating universe at the present day places a lower
bound on the coupling $W$ for a given $\lambda$ [see
Eq.~(\ref{lowerboundW})] and satisfying the
nucleosynthesis bound on $\lambda$ during the radiation era places a lower
bound on the value of the present day energy density of the scalar
field $\psi$:
\begin{equation}
\tilde{\Omega}_{\psi,0} \ge
\frac{1+\tilde{\Omega}_{\psi,nuc}^{Max}}{3},
\end{equation}
Thus a small energy density in the scalar field during the radiation
era, consistent with nucleosynthesis bounds, is compatible with a
large density in the present due to the different couplings of
relativistic and non-relativistic matter to the scalar field.

The scalar field behaves like a decaying cosmological constant whose
effective equation of state can be described by $w_q =
\tilde\gamma_\psi -1$, which is given from Eqs.~(\ref{defgammapsi})
and~(\ref{x5y5}) as
\begin{equation}
w_q = -\, {2W \over 2W-\sqrt{6}\lambda}
 = -\, {1\over2n-3} \ .
\end{equation}
The relationship between $\tilde\Omega_{M,0}=1-\tilde\Omega_{\psi,0}$
and $w_q$ is shown in Figure~1. In the limit of strong coupling
between the scalar field and the dark matter ($W\to\infty$ or
$\omega_{\rm DM}\to-3/2$) the equation of state approaches that of a
cosmological constant $w_q \to -1$. In contrast to the usual
cosmological models where a late time effective cosmological constant
is recovered only for a sufficiently flat potential, here $|\lambda|
\to \infty$ is allowed, whilst in the conformally related Brans-Dicke
frame $n$ remains finite [see Eq.~(\ref{deflambda})].  We see from
Figure~1 that if we require $\tilde\Omega_{M,0}\sim0.3$, then the
nucleosynthesis constraint on $|\lambda|$ drives $w_q$ very close to
$-1$.

\begin{figure}[t]
\centering 
\leavevmode\epsfysize=5cm \epsfbox{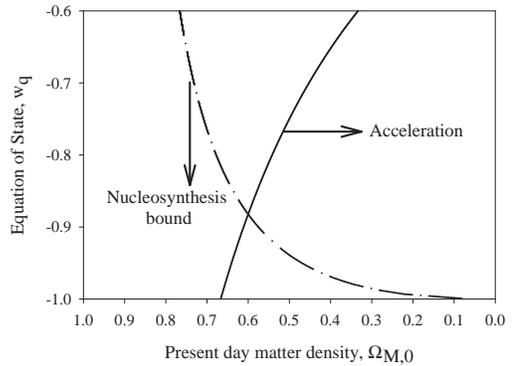}
\caption[Fig1] {$\Omega_{M,0}$-$w_q$ parameter space for 3-way
matter--kinetic--potential scaling solution with $\gamma=1$ and
$\lambda<0$.  Limits from nucleosynthesis constraints requiring
$\tilde{\Omega}_{\psi,{\rm nuc}}^{\rm Max}=0.2$ are shown by a
dot--dash line. The solid line shows the constraint from requiring
acceleration at present.} \label{Fig1}
\end{figure}

\section{Conclusions}
\setcounter{equation}{0}

We have presented the results of a phase-plane analysis of the
evolution of spatially flat FRW cosmological models containing a
barotropic fluid in a scalar-tensor theory of gravity where the
Brans-Dicke scalar, $\propto\phi^2$, has a self-interaction
potential, $V\propto\phi^{2n}$. In contrast to minimally coupled
fields in general relativity, this scalar field has self-similar
scaling solutions for an arbitrary power-law potential which
correspond to fixed points in the phase-plane.  The scalar-tensor
gravity theory with perfect fluid in the Brans-Dicke frame, is
equivalent to general relativistic gravity with an explicit
interaction between the fluid and the scalar field in the
conformally related Einstein frame. In this frame the effective
potential for the scalar field is an exponential and the
phase-plane coincides with that studied recently by
Amendola~\cite{Amendola99}, and Billyard and Coley~\cite{Bill99}.

There are five different possible fixed points for the system. Two of
these are solutions where the scalar field's kinetic energy dominates,
and the fixed points coincide with O'Hanlon and Tupper's vacuum
solutions in Brans-Dicke theory~\cite{OHT72}. A third solution, where
the scalar field's kinetic and potential energy remain proportional,
coincides with Mathiazhagan and Johri's~\cite{extinf} scalar-field
dominated power-law solution, which is inflationary for $\lambda^2<2$.
The fluid density is non-negligible only at the two remaining fixed
points. When the self-interaction potential energy of the scalar field
is negligible, we recover Nariai's power-law solutions for a
barotropic fluid in Brans-Dicke gravity. In these solutions, the rate
of change of the Brans-Dicke field, and hence the effective
gravitational constant $G\propto1/\phi^2$, is inversely proportional
to the Brans-Dicke parameter, $\omega$. But there is also a new
solution when the scalar field kinetic and potential energy densities
both remain proportional to the fluid density. In contrast to Nariai's
solutions, the rate of change of the effective gravitational constant
is independent of the value of $\omega$ and depends only upon the
potential exponent $n$.  This 3-way scaling solution was originally
discovered by in the context of Einstein
gravity~\cite{texas,Wetterich95}.

We have investigated whether it is possible in this theoretical
framework to construct cosmological solutions with non-relativistic
(pressureless) matter which are accelerating at late cosmic times.  A
generalised dark matter model may be one way to achieve this, where
non-relativistic dark matter is minimally coupled in a Brans-Dicke
frame, but baryonic matter is minimally coupled in the conformally
related Einstein frame (consistent with the current bounds on the
variability of the gravitational constant~\cite{Gue98}).  This leads
to an explicit energy transfer between the dark matter and the scalar
field in the Einstein frame. In contrast to other scalar field models
of quintessence~\cite{St98,Efstathiou99}, scalar fields with
potentials of pure exponential form, but non-minimally coupled to the
dark matter, allow a late-time scaling solution with equation of state
$w_q\approx-1$, as shown in Figure~1.  It is only possible for the
scalar field to make a significant contribution to the present energy
density, while remaining consistent with nucleosynthesis limits upon
the evolution during the radiation era, by approaching the strong
coupling limit ($\omega_{\rm DM}\to-3/2$).  Such strongly coupled dark
matter is very different from the usual model of non-interacting dark
matter and a more detailed investigation is necessary to determine
whether this could represent a viable cosmological model~\footnote
{Amendola~\cite{LucaII} has recently investigated large-scale structure
constraints on a coupled quintessence model.}.

Due to the scale-invariant nature of the theory and the existence of
self-similar cosmological solutions, there is no fine-tuning required
of the scalar field, either its initial conditions or its parameters,
in order to produce a change to an accelerated expansion at late
cosmic times. This can occur due to a change in the cosmological
equation of state when a non-relativistic dark matter component,
directly coupled to the scalar field, begins to make a significant
contribution to the total density. This is a familiar feature of the
standard hot big bang model where non-relativistic species, whether
baryons or massive neutrinos or more exotic dark matter candidates,
only come to affect the expansion at relatively late cosmic times.

\acknowledgments DJH acknowledges financial support from the
Defence Evaluation Research Agency.

\appendix
\section*{Appendix: Eigenvalues of critical points}
\renewcommand{\theequation}{A.\arabic{equation}}
\setcounter{equation}{0}

To determine the stability of solutions we expand about the critical
points $(x_i,y_i)$,
\begin{eqnarray}
x &=& x_i + u, \nonumber \\
y &=& y_i + v
\end{eqnarray}
which to first order gives the equations of motion for $u$ and $v$,
\begin{eqnarray}
u' & = & \left(\frac{9}{2}(2-\gamma)x_i^2 - 2Wx_i -
\frac{3}{2}\gamma y_i^2 -\frac{3}{2}(2-\gamma)\right) u \nonumber
\\ & + & \left((\lambda\sqrt{6} - 3\gamma x_i - 2W)y_i\right) v,
\nonumber
\\ v' & = & \left(3(2-\gamma)x_iy_i - \lambda \frac{3}{2} y_i
\right) u \nonumber \\ & + & \left (\frac{3}{2}(2-\gamma)x_i^2 -
\lambda\frac{3}{2}x_i - \frac{9}{2}\gamma y_i^2 +
\frac{3}{2}\gamma \right) v.
\end{eqnarray}
We solve the simultaneous first order equations for two eigenvalues
$m_\pm$ such that the eigenmodes $u=K_\pm v$ evolve as $u\propto
e^{m_\pm N}$.  Stability of the critical points $(x_i,y_i)$ depends on
the eigenvalues $m_\pm$. Points are stable if the real part of the
eigenvalues are negative, linear perturbations about the point being
exponentially damped. A saddle point in the phase plane has the real
part of the eigenvalues of opposite sign, whilst instability requires
both real parts to be positive.  The eigenvalues for the system are
listed below.

In what follows we will make extensive use of the following
bifurcation values for the parameters:
\begin{eqnarray}
\gamma_a &\equiv& 2 \left( 1 - {|W|\over3} \right) \ ,\\
\lambda_b &\equiv& \sqrt{6} \ ,\\
\sqrt{6}\lambda_c &\equiv& 2W + {9\gamma(2-\gamma) \over 2W} \ ,\\
\sqrt{6}\lambda_{d\pm} &\equiv& W \pm \sqrt{W^2+18\gamma} \ .
\end{eqnarray}

\begin{enumerate}

\item{2--Way, Matter--Kinetic scaling solution}
\begin{eqnarray}
m_- & = & - {3\over2}\left[ {(2-\gamma)^2-(2-\gamma_a)^2 \over
(2-\gamma)} \right]\,, \nonumber \\  m_+ & = & -
\frac{2W(\lambda-\lambda_c)}{\sqrt{6}(2-\gamma)}\,.
\end{eqnarray}
Exists for $\gamma<\gamma_a$ and stable for $W(\lambda-\lambda_c)>0$.

\item{Kinetic dominated solutions $x=\pm1, y=0$}
\begin{eqnarray}
m_+ & = & 3(2-\gamma) \mp 2W\,, \nonumber \\ m_- & = & 3 \left( 1
\mp {\lambda\over\sqrt{6}} \right)\,,
\end{eqnarray}
always exist. $x=\pm1$ stable for $\pm W>0$, $\gamma>\gamma_a$ and
$\pm\lambda>\lambda_b$.

\item{Scalar field dominated solutions}
\begin{eqnarray}
m_+ & = & - {1\over2} \left( 6 - \lambda^2 \right)\,, \nonumber \\
m_- & = & (\lambda-\lambda_{d-})(\lambda-\lambda_{d+})
\end{eqnarray}
Exists for $\lambda^2<\lambda_b^2$ and stable for
$\lambda_{d-}<\lambda<\lambda_{d+}$.

\item{3--Way, Matter--Kinetic--Potential scaling solutions}

Eigenvalues, $m_\pm$ are defined by the solution to the quadratic
equation,
\begin{equation}
m^2 + bm +c =0.
\end{equation}
where $b$ and $c$ are defined as,
\begin{eqnarray}
b = \frac{3}{(\sqrt{6}\lambda - 2W)^2} [
3(2-\gamma)(\lambda-\lambda_{d-})(\lambda-\lambda_{d+}) \nonumber
\\ \qquad -  2W\sqrt{6}(\lambda-\lambda_c)] \nonumber \,,
\end{eqnarray}
\begin{equation}
c = \frac{-6\sqrt{6} W (\lambda-\lambda_{d-})
(\lambda-\lambda_{d+}) (\lambda-\lambda_c)}
{(\sqrt{6}\lambda-2W)^2}\,,
\end{equation}
Exists for $W(\lambda_c-\lambda)>0$ and either $\lambda<\lambda_{d-}$
or $\lambda>\lambda_{d+}$. Always stable when it exists.

\end{enumerate}

\end{document}